\documentclass[11pt,twoside]{article}
\usepackage{asp2010}

\resetcounters

\begin{document}

\title{Unleashing the Power of Distributed CPU/GPU Architectures: Massive Astronomical Data Analysis and Visualization case study}
\author{A. H. Hassan$^1$, C. J. Fluke$^1$, and D. G. Barnes$^2$
\affil{$^1$ahassan@swin.edu.au, Centre for Astrophysics and Supercomputing, Swinburne University of Technology, POBox 218, Hawthorn, Australia, 3122}
\affil{$^2$Monash e-Research Centre, Monash University, Clayton, VIC 3800, Australia
}
}

\begin{abstract}
Upcoming and future astronomy research facilities will systematically generate terabyte-sized data sets moving astronomy into the Petascale data era. While such facilities will provide astronomers with unprecedented levels of accuracy and coverage, the increases in dataset size and dimensionality will pose serious computational challenges for many current astronomy data analysis and visualization tools. With such data sizes, even simple data analysis tasks (e.g. calculating a histogram or computing data minimum/maximum) may not be achievable without access to a supercomputing facility. 

To effectively handle such dataset sizes, which exceed today's single machine memory and processing limits, we present a framework that exploits the distributed power of GPUs and many-core CPUs, with a goal of providing data analysis and visualizing tasks as a service for astronomers. By mixing shared and distributed memory architectures, our framework effectively utilizes the underlying hardware infrastructure handling both batched and real-time data analysis and visualization tasks.  Offering such functionality as a service in a ``software as a service'' manner will reduce the total cost of ownership, provide an easy to use tool to the wider astronomical community, and enable a more optimized utilization of the underlying hardware infrastructure.

\end{abstract}

\section{Introduction}

Since they were first introduced for general purpose computing, graphics processing units (GPUs) have become a science-enabling technology across a wide variety of scientific fields [e.g. bioinformatics \citep{Michael_2007} and weather forecasting \citep{michalakes_2008}]. The lower cost per floating point operation, the low power consumption, and the sustainable speedup are all motivations to utilize GPUs as a practical high performance computing architecture - despite being somewhat harder to program than CPUs. Within the astronomical community, astronomers have adopted GPUs to approach many data processing and simulation problems [see \citet{I05_adassxxi}]. 

With the energy and the power consumption as a major obstacle toward further performance increase in current multi-core CPU computing architectures \citep{Kogge_2010}, it is anticipated that GPUs and other many-core architectures(e.g. field-programmable gate array and Cell processors) will be one of the main ways to address expected petascale data analysis and visualization problems. 
With datasets exceeding current single machine memory limits, and currently relatively low GPU memory (e.g. 6 GB\footnote{\url{http://www.nvidia.com/object/personal-supercomputing.html}}), it is vital to effectively address the problem of data handling and synchronization over heterogeneous distributed CPU/GPU architectures. 
  
Within this work, we are presenting a general purpose framework to effectively utilize heterogeneous multi-core CPUs and GPUs toward addressing data intensive high performance computing problems in astronomy. 

\section{Distributed GPU architecture}

\begin{figure}
\begin{center}
\includegraphics[scale=0.28]{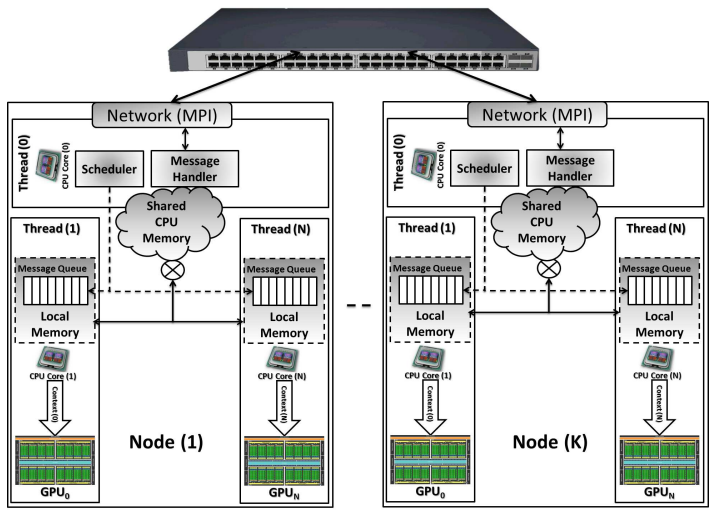}
\caption{Schematic diagram showing the main components of the framework. The framework is utilized to synchronize the communication between $K$ distributed nodes with $N$ GPUs each.}
\label{fig:MainDiag}                                 
\end{center}                                 
\end{figure}

Figure \ref{fig:MainDiag} shows the main framework components. Each GPU device within a node is managed through a CPU core, which is responsible for preparing the input data, invoking the GPU kernel in a synchronous manner, performing any necessary pre/post-processing, and sharing the data with the other threads.
 Each thread works as an independent process with a two-way communication with the master thread, which handles the communication between different threads (if needed) and the communication with the other nodes in a master-slave pattern.
 The communication is performed in an asynchronous manner between the master threads and other threads using a custom message queue at each thread. Different threads can access a shared memory space, allocated by the master threads, to share data and/or update its status, which is utilized by the scheduler sub-module for task allocation. 
 This access is controlled via one or more semaphores to ensure exclusive memory write. 
 Lately, GPU drivers have started to support the usage of a unified address space between GPU and CPU memory (e.g. NVIDIA CUDA 4.0\footnote{\url{http://www.nvidia.com/object/cuda_home_new.html}}), which can be utilized in this case as long as a control on the concurrent access to this shared memory is minimal or not required.\footnote{Atomic operations and concurrent access prevention usually degrade the GPU performance significantly.} 
 Another hardware feature which may be beneficial to speed-up data movement between different levels of the memory hierarchy is to use multiple execution queues (or streams) to overlap GPU computation with data I/O.      

All the communications between different distributed nodes are performed through the master threads only. Different data scattering and gathering operations are performed in two stages: a local stage between GPUs and CPUs using shared memory, and a global stage over the network using the message passing interface (MPI)\footnote{See \url{http://www.mcs.anl.gov/research/projects/mpi/} for details.} protocol. This partitioning, as long as it suits the problem, minimizes the amount of communication by a factor of $N$, where $N$ is the number of GPU units per node. 

To demonstrate the performance of the presented framework, we use interactive volume rendering of larger-than-memory spectral data cubes as a case study [see \citet{hassan_2011} for problem description and motivations]. 
With data exceeding single machine memory limits, real-time processing demands, and relatively high communication overhead that scales linearly with the number of processing elements, volume rendering (and interactive visualization in general) presents one of the worst case performance demands. It is a perfect example to demonstrate the power of mixing shared and distributed memory to achieve the highest possible performance.
 
The presented framework is developed as a server-side rendering back-end, with a remote visualization QT\footnote{\url{http://qt.nokia.com/products/}} desktop viewer to enable user interactivity and result display. The CUDA driver API was utilized to implement the GPU part, with MPI as the main communication software backbone between nodes.   

The performance in Table \ref{tab:resultstable} shows the framework performance [presented as number of frames per second (fps)] against the dataset size in GB for the same number of GPUs (128 GPUs)and processing nodes (64 nodes with 2 GPU each). The maximum achieved performance is 2.5 teravoxel processing per second. The amount of data exchanged is theoretically related to the output resolution which is megapixel/GPU. Due to different communication optimization and two-level gathering described before, the amount of data exchanged is reduced by at least 50\% \citep{hassan_2011}. The main distributed communication processing pattern was master-slave communication with no data compression. 

\begin{table*}
\caption{Performance output of the larger-than-memory volume rendering problem with different datasets ranging from 4 to 204 GB cubes over 128 GPUs and 64 nodes (2 GPUs per node).}
	\label{tab:resultstable}
	\centering
		\begin{tabular}{c|c|c|c}
		\hline
			Dimensions  & File Size & Tesla C1060 & Tesla C2050\\ 	
			{\scriptsize (Data Points)}&& {\scriptsize (240 cores and 4GB memory)} & {\scriptsize (448 cores and 3GB memory)}\\ \hline
			1024 x 1024 x 1024 &4 GB & 45 fps & 52 fps\\ \hline								
			2502 x 2501 x 1093 &26 GB & 41 fps & 52 fps  \\ \hline
			2600 x 2600 x 2600 &65 GB & 38 fps & 55 fps\\ \hline
			5004 x 5002 x 2186 &204 GB & 33 fps & 50 fps\\ \hline
		\end{tabular}	
\end{table*}

\section{Discussion}
The presented framework aims mainly to address the design and processing constraints of real-time problems, or problems which need different processing elements to communicate and exchange data in order to produce the final output results (e.g. global view data visualization or calculating the data median). 
This framework can address data exchange and synchronization demands of different data analysis tasks for datasets exceeding current single machine memory limits, especially when an in-situ data analysis is required to minimize I/O overhead. 
If we take, for example, an expected Australian Square Kilometre Array Pathfinder (ASKAP) spectral data cube (around 1TB), to do any processing on it using a single GPU would require partitioning the cube into 170 sub-cubes, with a different data loading for each one of them. This might be possible for a single-pass accumulative operation like calculating the data minimum and maximum, but cannot solve other multiple-pass problems such as calculating the median or standard deviation. 
More sophisticated data analysis tasks usually require the whole data to be in memory to perform measurement of global properties, and that is where our framework is more useful.

Addressing data analysis and visualization processes for such data volumes will need a clever resource utilization and data movement minimization to achieve reasonable computational performance. 
We think distributed GPUs can play a key role in enabling such tasks with reasonable response time. We showed in \citet{hassan_2011} that for a computationally intensive problem like volume rendering, replacing CPUs with GPUs as  the main processing element can dramatically reduce the number of processing nodes required. Consequently, this reduction decreases the communication overhead and the size of the computing facility required to address such problem.

Another aspect is working in a muti-user environment. With such data intensive problems we need a configurable, on-demand resource sharing model, which can fit our future needs [see \citet{Ostberg_2011} for a review of different available high performance computing resource management models]. 
We think the private cloud service oriented architecture may be a better resource sharing paradigm, with software, infrastructure and data offered as a service to the user via a remote thin-client. Preparing our software to integrate with such model can offer large scale distributed architecture in a more affordable way, reduce the total cost of ownership for both the software tools and infrastructure, and enhance access to large datasets.             

\acknowledgements A. Hassan thanks the Astronomical Society of Australia for their financial support.

\bibliographystyle{asp2010}
\bibliography{O19}

\end{document}